\documentclass[conference]{IEEEtran}
\usepackage[T1]{fontenc}
\usepackage[latin9]{inputenc}
\usepackage{color}
\usepackage{float}
\usepackage{amsthm}
\usepackage[nosumlimits]{amsmath}
\usepackage{amssymb}
\usepackage{multirow}
\usepackage{overpic}

\makeatletter

\floatstyle{ruled}
\newfloat{algorithm}{tbp}{loa}
\providecommand{\algorithmname}{Algorithm}
\floatname{algorithm}{\protect\algorithmname}

\theoremstyle{plain}

\theoremstyle{definition}

\theoremstyle{plain}

\theoremstyle{remark}

\theoremstyle{remark}

\allowdisplaybreaks

\usepackage{mathtools}

\usepackage{amsfonts}
\usepackage{cite}
\usepackage{array}
\usepackage{algorithm}
\usepackage{algorithmic}
\usepackage{graphicx}
\newcommand*{\rom}[1]{\expandafter\@slowromancap\romannumeral #1@}

\makeatother

\providecommand{\definitionname}{Definition}
\providecommand{\factname}{Fact}
\providecommand{\remarkname}{Remark}
\providecommand{\theoremname}{Theorem}
\providecommand{\lemmaname}{Lemma}

\mathchardef\mhyphen="2D

\usepackage{siunitx} 
\usepackage{subfig}
\usepackage[font = footnotesize]{caption}

\usepackage{url}

\begin{document}

\title{5G New Radio Evolution Towards Sub-THz Communications}

\author{
\IEEEauthorblockN{Oskari Tervo\textsuperscript{$\ast$}, Toni Levanen\textsuperscript{$\dag$}, Kari Pajukoski\textsuperscript{$\ast$}, Jari Hulkkonen\textsuperscript{$\ast$}, Pekka Wainio\textsuperscript{$\ast$}, and Mikko Valkama\textsuperscript{$\dag$} %
}
\IEEEauthorblockA{\textsuperscript{$\ast$} Nokia Bell Labs, Finland} 
\IEEEauthorblockA{\textsuperscript{$\dag$} Department of Electrical Engineering, Tampere University, Finland}
}

\maketitle



 \setlength{\belowdisplayskip}{5pt}
 \setlength{\abovedisplayskip}{5pt}
 \setlength{\textfloatsep}{3pt }

\begin{abstract}

In this paper, the potential of extending 5G New Radio physical layer solutions to support communications in sub-THz frequencies is studied.
More specifically, we introduce the status of third generation partnership project studies related to operation on frequencies beyond \SI{52.6}{GHz} and note also the recent proposal on spectrum horizons provided by federal communications commission (FCC) related to experimental licenses on \SI{95}{GHz} - \SI{3}{THz} frequency band. Then, we review the power amplifier (PA) efficiency and output power challenge together with the increased phase noise (PN) distortion effect in terms of the supported waveforms. As a practical example on the waveform and numerology design from the perspective of the PN robustness, link performance results using \SI{90}{GHz} carrier frequency are provided. The numerical results demonstrate that new, higher subcarrier spacings are required to support high throughput, which requires larger changes in the physical layer design. It is also observed that new phase-tracking reference signal designs are required to make the system robust against PN. The results illustrate that single-carrier frequency division multiple access is significantly more robust against PN and can provide clearly larger PA output power than cyclic-prefix orthogonal frequency division multiplexing, and is therefore a highly potential waveform for sub-THz communications.

\end{abstract}


\section{Introduction}

5G New Radio (NR) standards release (Rel) 15 and Rel-16 support carrier frequencies up to \SI{52.6}{GHz}. As an initial effort to enable and optimize 5G NR system for operation in above \SI{52.6}{GHz}, third generation partnership project (3GPP) radio access network (RAN) specification group has studied requirements for 5G NR operation in frequency band \SI{52.6}{GHz} - \SI{114.25}{GHz} \cite{3GPPTR38807}. This study included global spectrum availability and regulatory requirements, potential use cases and deployment scenarios, and 5G NR system design requirements and considerations. The frequencies above \SI{52.6}{GHz} are faced with more difficult challenges, such as higher phase noise (PN), extreme propagation loss, high atmospheric absorption in certain frequencies, lower power amplifier (PA) efficiency, and strict transmitted power spectral density regulatory requirements, when compared to lower frequency bands. The interest in the higher carrier frequencies is driven by the fact that the frequency ranges above \SI{52.6}{GHz} potentially contain larger spectrum allocations and larger bandwidths that are not available below \SI{52.6}{GHz}. The 3GPP has recently decided to start further Rel-17 study and work items for system design aspects for 5G NR beyond \SI{52.6}{GHz} operation, where the main interest will be to first extend the current NR frequency range 2 (FR2) support to the frequency range \SI{52.6}{GHz} - \SI{71}{GHz} with minimal changes to the system \cite{RP-193259,RP-193229}. This is to make use of the existing commercial potential in the spectrum between \SI{57}{GHz} and \SI{71}{GHz}.

 \begin{figure*}
     \centering
     \includegraphics[angle=0,width=1.8\columnwidth]{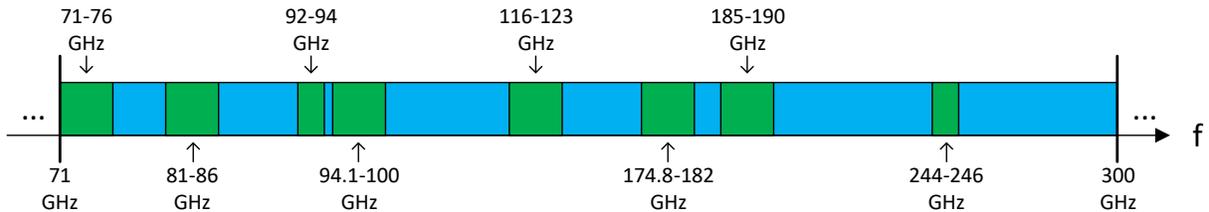}
     \caption{Example spectrum allocations (denoted with green color) between \SI{71}{GHz} and \SI{300}{GHz} based on current FCC regulation.}
     \label{fig:frequencyBandExample}
 \end{figure*}

However, it is expected that future studies will continue to beyond \SI{71}{GHz} and even higher sub-THz frequencies, and the current specification optimized for below \SI{52.6}{GHz} has to be significantly revised. This is clear already due to the fact that the transceiver impairments, such as PN, will increase drastically \cite{2018:B:Dahlman:5GNR}. To this end, the supported waveforms and subcarrier spacings (SCSs) will be some of the first design aspects of the system that need to be revised. Currently supported waveforms in 5G NR are orthogonal frequency division multiplexing (OFDM) and single-carrier frequency division multiple access (SC-FDMA). The latter is currently only supported for uplink and rank-1 transmissions, because it is mainly designed for coverage limited cases due to its better power efficiency. The rank of the transmission defines the number of simultaneously transmitted spatial layers, relating to traditional spatial multiplexing in multi-antenna systems.

Current spectrum availability is still limited in the frequencies beyond \SI{52.6}{GHz} \cite{3GPPTR38807}. Globally, there is an unlicensed spectrum at frequency range \SI{59}{GHz} - \SI{64}{GHz} but, e.g., in USA, the whole \SI{57}{GHz}-\SI{71}{GHz} frequency range is available, providing \SI{14}{GHz} of continuous bandwidth. In addition, the \SI{66}{GHz} - \SI{71}{GHz} band was just recently opened for licensed use in international telecommunications union (ITU) Regions 1 and 3 \cite{WRC-19}. The frequencies above \SI{71}{GHz} are mainly reserved for fixed, point-to-point communications, except in USA where both mobile and fixed communications are allowed. Considering the regulation in Europe for frequency band \SI{71}{GHz}-\SI{100}{GHz}, up to \SI{18}{GHz} aggregated bandwidth is available for fixed communications. 
To expedite the development towards sub-THz frequencies, FCC has decided to adopt new rules for the bands above \SI{95}{GHz} \cite{FCC:SpectrumHorizons}. This means that innovators can get experimental licenses, so called Spectrum Horizons Licences, to use frequencies from \SI{95}{GHz} up to \SI{3}{THz}, including also \SI{21.2}{GHz} of spectrum for unlicensed devices. The potential spectrum allocations for sub-THz communications based on current FCC regulation are illustrated in Fig. \ref{fig:frequencyBandExample}.

 \begin{figure}
     \centering
     \includegraphics[angle=0,width=0.99\columnwidth]{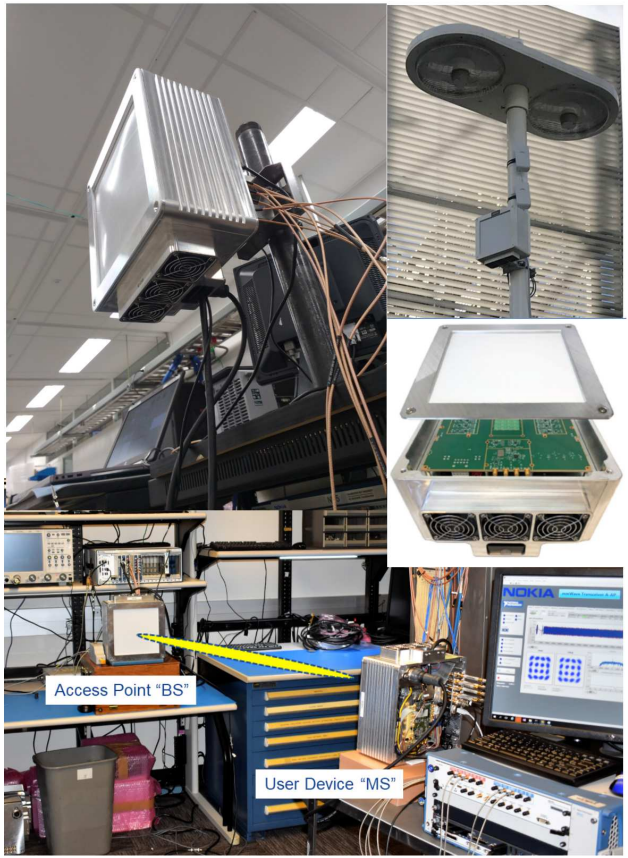}
     \caption{Proof-of-Concept system \cite{Iwabuchi-19}: the \SI{93}{GHz} radio head and the access point - user device test link.}
     \label{fig:TestBed}
 \end{figure}

As an example of practical development towards sub-THz frequencies, experimental mobile communication systems using frequency bands above \SI{52.6}{GHz} have been recently reported with performance evaluation results in both indoor and outdoor environments \cite{Inoue-17,Sanchis-Borras-17,Iwabuchi-19}. Especially in \cite{Iwabuchi-19}, experiments with a Proof-of-Concept demonstrator system operating at \SI{93}{GHz} band are discussed. The presented system, illustrated also in Fig. \ref{fig:TestBed}, is using a large 384 element phased-array antenna based on an advanced W-band (\SI{75}{GHz}-\SI{110}{GHz}) radio frequency integrated circuit (RFIC) described in \cite{Shahramian-18,Shahramian-19}, and advanced beam tracking techniques to support high data rates and mobility for evolved 5G NR systems. In this experiment the used single carrier waveform and numerology followed the principles defined in \cite{Inoue-17}.

To recognize the potential problems and solutions of extending the current 5G NR to higher frequencies, this paper studies physical layer numerology and waveforms for
sub-THz communications. We will focus on the OFDM and SC-FDMA waveforms currently supported in 5G NR Rel-15. First the significant difference in the achievable PA output power with these waveforms is demonstrated, showing the importance of supporting SC-FDMA in uplink and downlink. Then, the performance of OFDM and SC-FDMA under PN is investigated, using the highest currently supported SCS, and also various higher SCS options, to find out the required numerology to enable high throughput communications in sub-THz frequencies.

An extensive set of performance results assuming either rank-1 or rank-2 transmission at \SI{90}{GHz} carrier frequency are provided. In the performance evaluations, we assume the 5G NR Rel-15 based phase tracking reference signal (PTRS) designs as a reference, and illustrate the potential of using enhanced PTRS configurations to improve the performance with both evaluated waveforms. 
The numerical results indicate that the straightforward extension of the current 5G NR Rel-15 system to sub-THz bands is not possible. The first observation is that increased SCS is required, and, thus, larger changes for physical layer design are also required. Secondly, new PTRS designs are required to handle the PN more effectively, especially when multi-rank transmission is used. In general, the results demonstrate that SC-FDMA can provide significant improvement in the PA efficiency and larger PA output powers, and it is more robust under PN, and thus, it should be considered as the main waveform candidate for downlink and uplink communications.

\section{Design Principles for sub-THz Operation}
\label{sec:5GNowAndBeyond71GHz}

\subsection{Physical Layer Numerology}

The 5G NR is designed to support wide range of SCSs to handle different use cases and a wide range of supported carrier frequencies. Rel-15 defines two separate frequency ranges with different numerologies \cite{2017:J:Parkvall:5GNR}. The frequency range 1 (FR1) is defined for carrier frequencies \SI{410}{MHz} - \SI{7.125}{GHz} and supports SCSs 15/30/60~kHz, while FR2 is currently defined for frequency range \SI{24.25}{GHz} - \SI{52.6}{GHz} and supports 60/120/240~kHz SCSs. The largest SCS \SI{240}{kHz} is only allowed for the synchronization signal block to reduce the symbol length for beam sweeping procedure \cite{3GPPTS38300}. 

The physical layer numerology has many effects on the system. If one increases the SCS, the bandwidth and symbol rate increases for a given FFT size, and thus faster processing is required in the transceiver. Increased symbol rate is desired for narrow beam based operation, because beam training can be performed faster. Larger SCS is also more robust to inter-carrier interference (ICI) caused by PN. The time duration of the cyclic prefix (CP) decreases with increasing SCS for the same CP overhead. This may not be a problem from the inter-symbol interference point-of-view, because narrow beams result in very small delay spreads \cite{2015:C:Kim-directionalDelaySpread}. However, too short CP may become a problem for the beam-based transmission, if the beam switching delay is assumed to be comparable to CP length. Furthermore, increasing the SCS may have a larger impact on the system design as, e.g., the scheduling periods become too short and control channel coverage can be degraded. On the other hand, use of lower SCSs together with carrier aggregation would enable better use of legacy systems.

In this work, we assume that the supported SCSs in sub-THz communications could correspond to 120/240/480/960/1920/3840~kHz. The reasoning for this assumption is to follow the Rel-15 scaling principle from \SI{15}{kHz} SCS with powers of two \cite{2017:J:Parkvall:5GNR}, enabling easy and flexible frequency-domain and time-domain alignment of different numerologies.

\subsection{Power Amplifier Efficiency and Output Back-off}

Higher frequencies have many limitations which require careful study. The first major drawback requiring consideration is the decreased PA efficiency at higher carrier frequencies. For example, in \cite[Section 6.1.9.1]{3GPPTR38803}, it is shown that the output power of PAs for a given integrated circuit technology roughly degrades by \SI{20}{dB} per decade. This imposes a significant need to support waveforms and modulations that allow to achieve very low peak-to-average-power-ratio (PAPR) in order to achieve better power efficiency in base station (BS) and user equipment (UE) side, and to achieve the targeted maximum transmitted power levels. Maximizing the transmitter power is important because it directly translates into maximizing the cell coverage. It is well known that OFDM signal has larger PAPR than SC-FDMA \cite{2013:B:Dahlman:4GLTE}, especially at lower modulation orders, which emphasizes the importance of supporting SC-FDMA in downlink and uplink for sub-THz communications.

\begin{figure}
     \centering
     \includegraphics[angle=0,width=0.99\columnwidth]{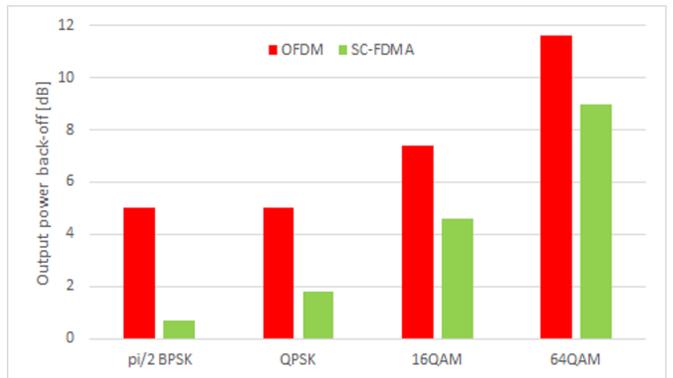}
     \caption{Comparison of PA output power back-off with \SI{20}{dB} ACLR requirement and modulation specific EVM requirement.}
     \label{fig:MPR}
 \end{figure}

In Fig. \ref{fig:MPR}, the required output back-off is shown for OFDM and SC-FDMA waveforms, when assuming \SI{20}{dB} adjacent channel leakage ratio (ACLR) requirement and modulation specific error vector magnitude (EVM) requirement. It is observed that depending on the modulation order, OFDM requires approximately from \SI{3}{dB} to \SI{5}{dB} more output power back-off, indicating that SC-FDMA is able to provide significantly better coverage.

\subsection{Phase Noise and PTRS Designs}

Another major impairments is the PN, which increases \SI{6}{dB} for every doubling of the carrier frequency \cite{2018:B:Dahlman:5GNR}. For example, going from \SI{28}{GHz} to \SI{90}{GHz} causes a ten-fold increase of PN. Thus, the maximum currently supported SCS of \SI{120}{kHz} for data transmissions may be not enough. Since OFDM and SC-FDMA use multiple orthogonal subcarriers to transmit data, they are affected quite similarly by PN distortion. More specifically, PN causes common phase error (CPE) which affects all subcarriers similarly \cite{2019:L:Syrjala:blockPTRS}. This distortion can be simply compensated in the receiver by derotating all subcarriers with the estimated CPE value. However, the significantly higher PN in the considered frequency range will cause also ICI, which comes from the convolution of the PN frequency response with the data bearing subcarriers. The degrading effect of ICI can be reduced or mitigated by increasing the SCS, or by applying PTRS designs which allow the estimation and compensation of the ICI components. The currently supported waveforms OFDM and SC-FDMA perform differently under different PTRS designs and related PN mitigation techniques and one of the main design challenges is the choice of the waveform and numerology, together with effective PN compensation methods.  

Current 5G NR handles the PN distortion by exploiting PTRSs, which are inserted into OFDM or SC-FDMA symbols to be able to track the PN variations. The OFDM uses distributed frequency-domain PTRSs, which enable the receiver to compensate only common phase error part of the PN. This leads to degraded performance with 5G NR Rel-15 numerology when considering sub-THz communications, as will be shown in Section \ref{sec:results}. The PTRS symbols are carried by individual subcarriers in every second or fourth physical resource block (PRB) and in every, every second, or every fourth OFDM symbol \cite[Section 7.4.1.2]{3GPPTS38211}.  

On the other hand, SC-FDMA uses time-domain PTRSs, which allow the receiver to track also the time varying nature of the PN within the SC-FDMA symbol. The PTRS symbols are grouped into groups of two or four time domain sub-symbols within the SC-FDMA symbol, and there can be two, four, or eight PTRS groups within the SC-FDMA symbol \cite[Section  6.4.1.2]{3GPPTS38211}. With SC-FDMA, the PTRS design allows a computationally efficient implementation to track and compensate time-varying PN response within a SC-FDMA symbol, which is not possible with the Rel-15 NR distributed PTRS design for OFDM. 

To enable PN induced ICI estimation with OFDM, we use the concept of frequency domain block PTRS introduced in \cite{2019:L:Syrjala:blockPTRS}. The basic idea is to allocate a frequency contiguous block of PTRS symbols which allows to estimate frequency-domain ICI components at the receiver. Block PTRS is inserted to each OFDM symbol, as the time continuity of ICI components is typically not guaranteed. 
In addition, as the current Rel-15 NR specification dictates a specific frequency resolution for distributed PTRS, it is possible that with block PTRS based design one can achieve better performance with lower reference signal overhead in wide channels, thus making it especially suitable for sub-THz communications. 

In order to improve the PN estimation capability with SC-FDMA, we propose to use new PTRS configurations which allow to increase the number of PTRS groups per SC-FDMA symbol to improve the receiver capability to track fastly changing PN time response. More detailed description of the Rel-15 and enhanced PTRS designs can be found from \cite{2019:J:Levanen-B50GHz}, including a comprehensive evaluation of the benefits of these methods related to 5G NR operation in the \SI{60}{GHz} carrier frequency.

\section{Radio Link Performance at \SI{90}{GHz} Frequency}
\label{sec:results}

{
\floatstyle{boxed} 
\restylefloat{figure}
\begin{figure*}[t]
  \centering
  \subfloat[][]{\includegraphics[angle=0,width=0.85\columnwidth]{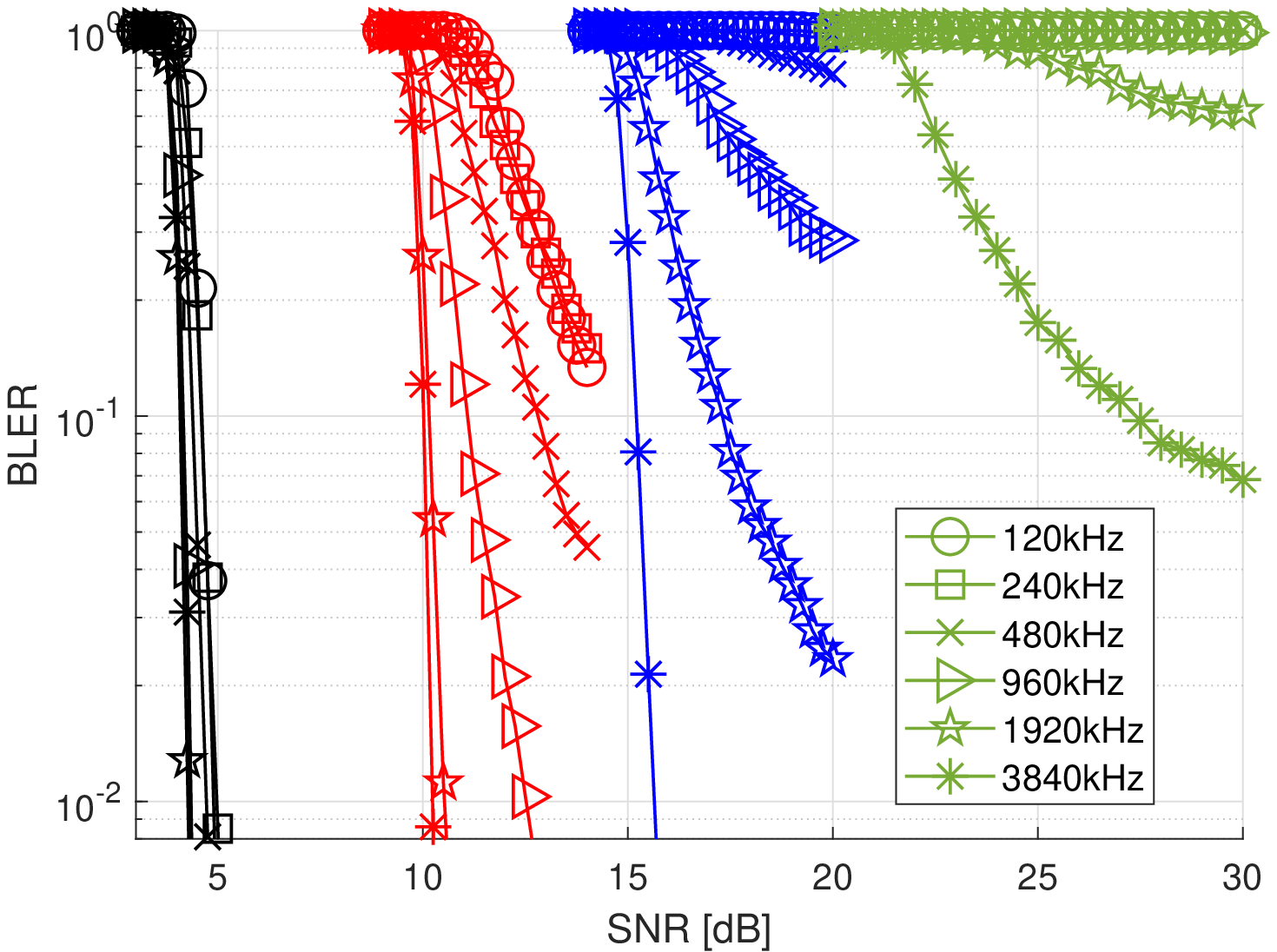}}
  \qquad
  \subfloat[][]{\includegraphics[angle=0,width=0.85\columnwidth]{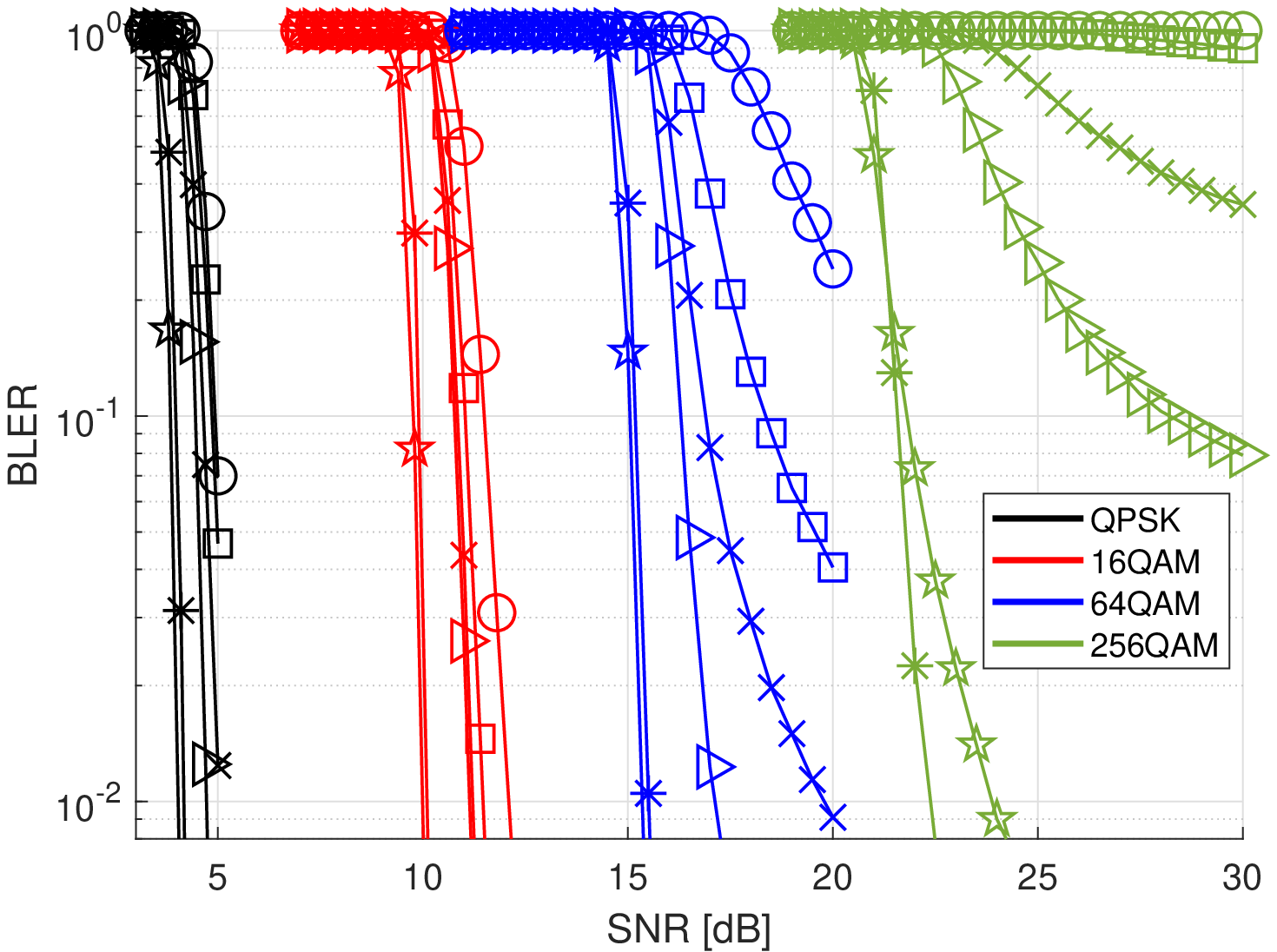}}
  \vskip\baselineskip
  \subfloat[][]{\includegraphics[angle=0,width=0.85\columnwidth]{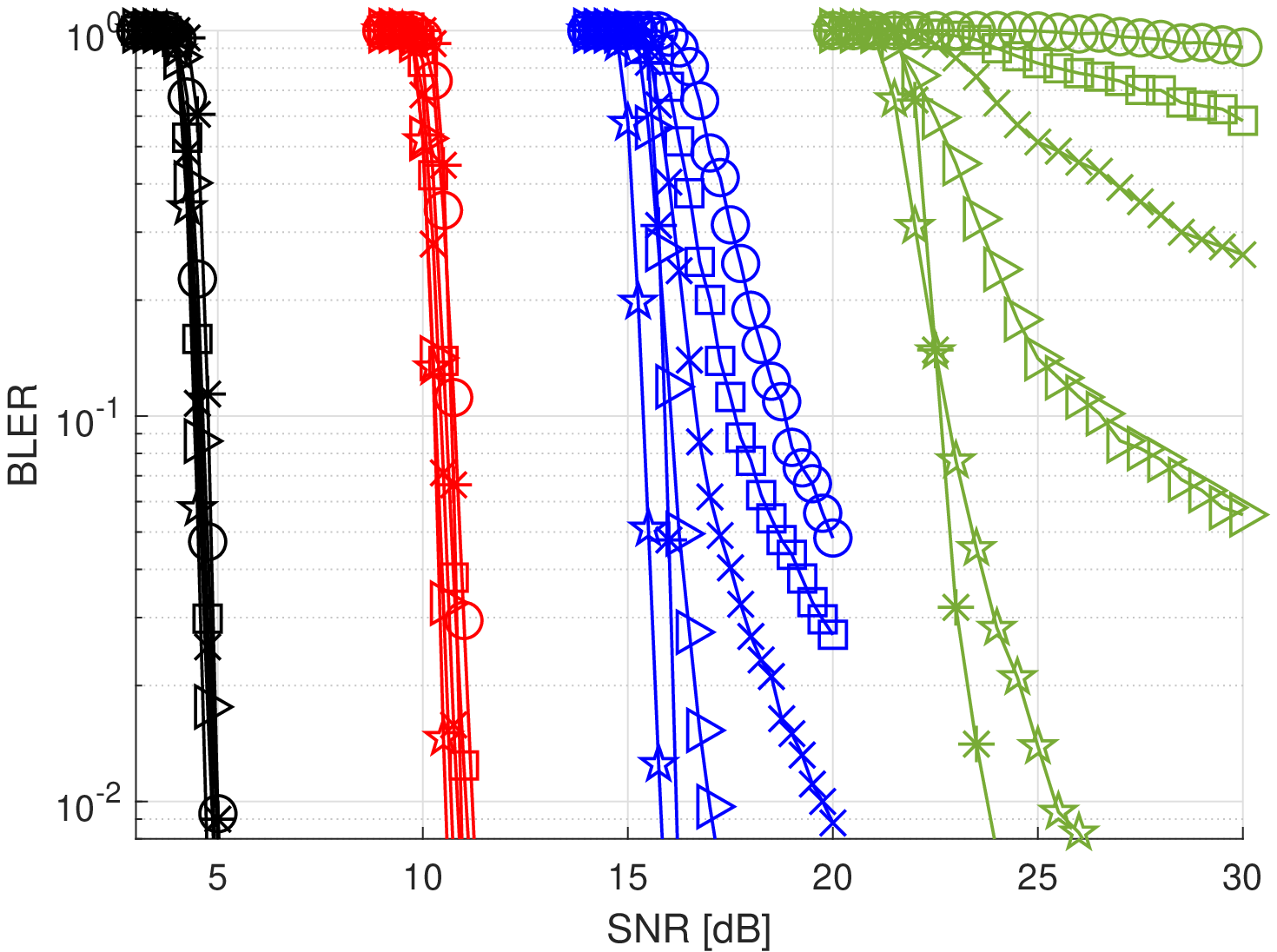}}
  \qquad
  \subfloat[][]{\includegraphics[angle=0,width=0.85\columnwidth]{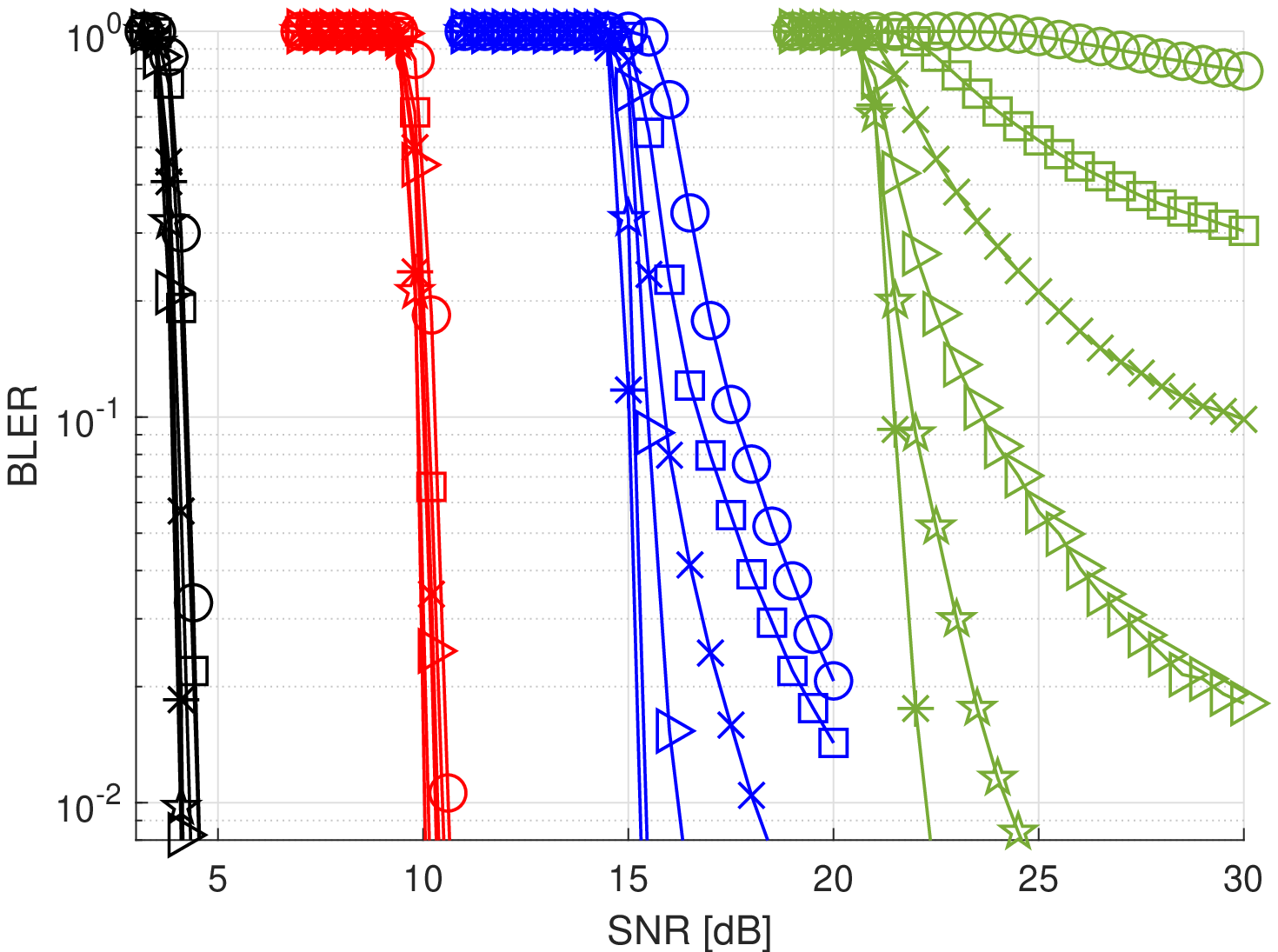}}
   \caption{Link performance with PN for a) OFDM and b) SC-FDMA using Rel-15 compliant PTRS and for c) OFDM and d) SC-FDMA using enhanced PTRS structures. The carrier frequency is \SI{90}{GHz} and rank-2 transmission is assumed. }
  \label{fig:90GHz_WithPTRS}
\end{figure*}
}

In this section, the performance of OFDM and SC-FDMA waveforms using PTRSs is evaluated over different SCSs. We assume a maximum channel bandwidth of \SI{2.16}{GHz}, which follows the channelization for WLAN 802.11ay operating in \SI{60}{GHz} unlicensed band \cite{2018:J:Zhou:80211ayTutorial}. Therefore, the maximum allocation size is limited to 180, 90, or 45 PRBs with SCS \SI{960}{kHz}, \SI{1920}{kHz}, or \SI{3840}{kHz}, respectively. For the smaller SCS, we have limited the maximum allocation size to 180 PRBs also with SCSs less than \SI{960}{kHz}. This is to keep the transmission block size constant with 120-\SI{960}{kHz} SCS, so that we can fairly see the impact of phase noise to the performance, e.g., by eliminating the effect of rate matching with varying allocation size. Another reason is that one can basically use carrier aggregation for \SI{120}{kHz} SCS to get exactly the same bandwidth as with \SI{960}{kHz} SCS.

There are 128 antenna elements organized into a 8x16 antenna array per polarization at the BS, and 16 antenna elements organized into a 4x4 antenna array per polarization at the UE. 
With this architecture, the maximum supported transmission rank is two,  
which is thus in the focus of the evaluations. In addition, rank-2 transmission with a given modulation order achieves a specific throughput with smaller SNR than rank-1 with doubled modulation order, as will be observed in Table \ref{tab:rank1rank2}, thus making it highly efficient scheme for increased spectral efficiency operation in sub-THz communications. The link performance is evaluated with QPSK, 16-QAM, 64-QAM, and 256-QAM modulations with fixed coding rate of $R=2/3$, and the used channel codec is a 5G Rel-15 NR compliant LDPC code. The used channel model is clustered delay line E (CDL-E) with \SI{10}{ns} root-mean-squared (RMS) delay spread and Rician factor $K=$\SI{15}{dB} \cite{3GPPTR38901}. In all cases, a UE mobility of \SI{3}{km/h} is assumed. 

We use the PN models defined in \cite[Section 6.1.11]{3GPPTR38803}, which are different for BS and UE side, so that PN is higher at the UE side due to stricter power consumption requirement. Both antenna ports are assumed to experience the same PN, which models the case where the same local oscillator signal is shared between the antenna ports. With this assumption, the PTRSs can be inserted only to one antenna port. 

For the 5G NR Rel-15 references, we have assumed the maximum density for PTRS, for both OFDM and SC-FDMA, to achieve the best possible performance. In the case of OFDM with block PTRS \cite{2019:L:Syrjala:blockPTRS}, a contiguous allocation of 4~PRBs is used. On the RX side four frequency components from both sides of the zero-frequency (DC) component of the PN frequency response are estimated, 
including the DC component, and used in the PN compensation. With enhanced PTRS configuration for SC-FDMA, a design leading to the same overhead as the block PTRS proposed for OFDM waveform was selected. Thus, a new PTRS configuration using 12 PTRS groups with four PTRS sub-symbols per group is used. 

The performance of OFDM using either Rel-15 distributed PTRS design or block PTRS design is shown in Figures \ref{fig:90GHz_WithPTRS} (a) or (c), respectively, together with SC-FDMA following either Rel-15 based or enhanced PTRS design in Figures \ref{fig:90GHz_WithPTRS} (b) or (d), respectively. 

Fig. \ref{fig:90GHz_WithPTRS} (a) shows that Rel-15 PTRS design for OFDM cannot support even 64-QAM if SCS smaller than $\SI{1920}{kHz}$ is used. This already reveals that the currently supported highest SCS \SI{120}{kHz} cannot be used for OFDM with distributed PTRS, but instead a significant increase of SCS is required for sub-THz communications. On the other hand, as shown in Fig. \ref{fig:90GHz_WithPTRS} (b), SC-FDMA with Rel-15 PTRS design performs significantly better with 64-QAM, so that even \SI{240}{kHz} SCS, which is already supported for beam training in Rel-15, could be used with limited performance loss. From Fig. \ref{fig:90GHz_WithPTRS} (a), we also note that OFDM with Rel-15 PTRS cannot basically support 256-QAM at all, whereas SC-FDMA with Rel-15 PTRS can support it with \SI{1920}{kHz} SCS. Nevertheless, it is observed that Rel-15 physical layer cannot be used as such for communications in \SI{90}{GHz} carrier frequency or beyond. Without modifying the PTRS designs, the only option to support higher-order modulations would be to use SC-FDMA together with significantly increased SCS.

Fig. \ref{fig:90GHz_WithPTRS} (c), shows that using block PTRS with OFDM improves the performance significantly when compared to Rel-15 PTRS results shown in Fig. \ref{fig:90GHz_WithPTRS} (a). We can see that using block PTRS, 64-QAM could be supported with existing \SI{120}{kHz} SCS, giving approximately \SI{4}{dB} performance degradation. However, it is observed that supporting 256-QAM would still require increasing the maximum SCS to \SI{960}{kHz}. When looking at SC-FDMA results with enhanced PTRS configuration in Fig. \ref{fig:90GHz_WithPTRS} (d), a clear improvement compared to Rel-15 performance is observed. Now all SCSs provide similar performance with QPSK and 16-QAM modulations. With 64-QAM modulation, the degradation of using \SI{120}{kHz} SCS is only approximately \SI{3}{dB}, indicating that \SI{120}{kHz} SCS could still be used for communications in \SI{90}{GHz} carrier frequency. For 256-QAM modulation, a clear improvement in the link performance with \SI{960}{kHz} SCS is observed, allowing it to be used even for the extreme throughput scenario. In general, with enhanced PTRS designs, SC-FDMA provides slightly better link performance than OFDM in all cases with equal PTRS overhead. These results highlight that in order to enable very high data rates using 256-QAM, at least \SI{960}{kHz} SCS is required, and that SC-FDMA performs significantly better in all cases when compared to OFDM. 

Table \ref{tab:rank1rank2} shows the performance difference between rank-1 and rank-2 transmissions in terms of required signal-to-noise-ratio (SNR) to reach 10\% BLER target, when using enhanced PTRS designs. We can first observe the benefit of using rank-2 transmission. As an example, QPSK with rank-2 requires about \SI{4}{dB} SNR to achieve the same throughput as 16-QAM with rank-1, which requires about 7dB SNR. Similarly 16-QAM with rank-2 is drastically better than 256-QAM with rank-1. Noting the PA efficiency and output power achieved with SC-FDMA and the enhanced spectral efficiency with rank-2 transmission, it is clear that rank-2 transmission with lower order modulations is highly suitable for sub-THz communications. 

The results in Table \ref{tab:rank1rank2} also indicate that rank-2 can be used with QPSK and 16-QAM with approximately \SI{3}{dB} higher SNR than rank-1 for all evaluated SCSs, which is a fundamental lower limit for the performance difference. For 64-QAM, we start to see increased performance loss when SCS is smaller than \SI{480}{kHz}. However, these smaller SCSs can be still used and the SNR gap could be reduced by further optimizing the PTRS designs. However, with 256-QAM, at least \SI{1920}{kHz} SCS is required to efficiently support rank-2 transmissions, and \SI{960}{kHz} could be used by further optimizing the PTRS designs. 

We can conclude that clear increase of the SCS is required to enable extreme data rates, especially for rank-2 transmissions. However, lower SCSs can be still used with lower order modulations to extend the coverage of the system.

\begin{table}[]
    \caption{Comparison of the required SNR (in dB) for reaching 10\% BLER target, assuming rank-1 or rank-2 spatial multiplexing and using the enhanced PTRS designs. In the table, N/A means that the 10\% BLER target cannot be achieved in the evaluated SNR regime, and column "Diff" denotes the difference in the required SNR between rank-1 and rank-2 transmissions.}
    \centering
    \begin{tabular}{|c|c|c|c|c|c|c|}
        \hline
        & \multicolumn{3}{|c|}{SC-FDMA} & \multicolumn{3}{|c|}{OFDM} \\ \hline SCS [kHz] & rank-1 & rank-2
         & Diff. & rank-1 & rank-2 & Diff.
        \\
        \hline &
        \multicolumn{6}{c|}{\textbf{QPSK}} \\ \hline
        120 & 1.2 & 4.2 & 3.0 & 1.3 & 4.6 & 3.3
        \\ \hline
         240 & 1.2 & 4.2 & 3.0 & 1.2 & 4.6 & 3.4
        \\ \hline
         480 & 1.0 & 4.0 & 3.0 & 1.2 & 4.5 & 3.3
         \\ \hline
         960 & 0.9 & 3.9 & 3.0 & 1.2 & 4.5 & 3.3
         \\ \hline
         1920 & 0.9 & 3.9 & 3.0 & 1.2 & 4.4 & 3.2
         \\ \hline
         3840 & 0.8 & 3.9 & 3.1 & 1.2 & 4.8 & 3.6
        \\
        \hline &
        \multicolumn{6}{c|}{\textbf{16-QAM}} \\ \hline
        120 & 7.2 & 10.3 & 3.1 & 7.2 & 10.8 & 3.1
        \\ \hline
         240 & 7.0 & 10.1 & 3.1 & 7.0 & 10.6 & 3.1
        \\ \hline
         480 & 7.0 & 10.0 & 3.0 & 7.0 & 10.4 & 3.0
         \\ \hline
         960 & 7.0 & 10.0 & 3.0 & 7.0 & 10.3 & 3.0
         \\ \hline
         1920 & 6.8 & 9.9 & 3.1 & 6.9 & 10.3 & 3.1
         \\ \hline
         3840 & 6.8 & 9.9 & 3.1 & 7.0 & 10.7 & 3.1      
        \\
        \hline &
        \multicolumn{6}{c|}{\textbf{64-QAM}} \\ \hline
        120 & 13.4 & 17.6 & 4.2 & 13.9 & 18.8 & 4.9
        \\ \hline
         240 & 12.7 & 16.7 & 4.0 & 13.2 & 17.6 & 4.4
        \\ \hline
         480 & 12.4 & 15.9 & 3.5 & 12.8 & 16.7 & 3.9
         \\ \hline
         960 & 12.3 & 15.5 & 3.2 & 12.5 & 16.0 & 3.5
         \\ \hline
         1920 & 12.1 & 15.2 & 3.1 & 12.0 & 15.9 & 3.9
         \\ \hline
         3840 & 11.9 & 15.0 & 3.1 & 11.9 & 15.4 & 3.5                  \\
        \hline &
        \multicolumn{6}{c|}{\textbf{256-QAM}} \\ \hline
        120 & N/A & N/A & N/A & N/A & N/A & N/A
        \\ \hline
         240 & 23.2 & N/A & N/A & N/A & N/A & N/A
        \\ \hline
         480 & 20.0 & 29.8 & 9.8 & 23.6 & N/A & N/A
         \\ \hline
         960 & 18.2 & 23.6 & 5.4 & 20.0 & 26.5 & 6.5
         \\ \hline
         1920 & 18.2 & 21.9 & 3.7 & 18.4 & 22.8 & 4.4
         \\ \hline
         3840 & 18.2 & 21.5 & 3.3 & 18.2 & 22.6 & 4.4 
         \\ 
        \hline
    \end{tabular}
    \label{tab:rank1rank2}
\end{table}

\section{Conclusions}
\label{sec:conclusions}

In this paper, the numerology and waveform evolution of 5G NR towards sub-THz communications has been analyzed. The main implementation problems and available bands for mobile communications have been discussed. The results demonstrate that SC-FDMA would be significantly better waveform choice due to the following reasons. First, it provides consistently better link performance under phase noise with only minor changes in the Rel-15 PTRS design. Secondly, SC-FDMA waveform can enable clearly better coverage, because it provides significantly larger power amplifier output power than OFDM, especially with low-order modulations. 

The extensive set of provided results indicate that direct extension of 5G NR Rel-15 physical layer to sub-THz communications is only possible with low-order modulations due to the phase noise impact. With \SI{120}{kHz} SCS and Rel-15 PTRS designs, OFDM supports 16-QAM and SC-FDMA could support 64-QAM with significant SNR degradation. With the enhanced PTRS designs and currently supported \SI{120}{kHz} SCS, OFDM and SC-FDMA could enable modulations up to 64-QAM with reduced SNR degradation. The results indicate that to support 64-QAM or 256-QAM without significant SNR degradation, significant increase in the supported SCS is required, especially for rank-2 transmissions. The lower-order modulations can support all SCSs, including \SI{120}{kHz} SCS, to allow extended coverage operation in the sub-THz communications system. Furthermore, enhanced PTRS designs supporting rank-2 transmission and 64-QAM with \SI{120}{kHz} SCS allow to use currently available 5G NR FR2 solutions in sub-THz communications.

Although the 3GPP Rel-17 studies are currently limited to \SI{71}{GHz} carrier frequency, the research is evolving towards 6G, which is expected to bring even higher, sub-THz frequencies into consideration. This paper demonstrates that physical layer requires rethinking when going to very high frequencies, in terms of supported subcarrier spacings and downlink waveforms, which can have a significant impact on the whole physical layer design.

\bibliographystyle{IEEEtran}
\bibliography{IEEEabrv,references}

\end{document}